 \documentclass[smallabstract,smallcaptions]{dccpaper}

\usepackage{epsfig}
\usepackage{citesort}
\usepackage{amsmath}
\usepackage{amssymb}
\usepackage{color}
\usepackage{url}
\usepackage{booktabs}
\usepackage{subfigure}
\usepackage{multirow}

\newlength{\figurewidth}
\newlength{\smallfigurewidth}

\newcommand{\getCellOpName}[0]{\mathit{access}}

\newcommand{\getCellOpNew}[3]{\mathit{access}(#1,#2,#3)}
\newcommand{\rangeQueryOpName}[0]{\mathit{windowQuery}}

\newcommand{\rangeQueryOpNew}[6]{\mathit{windowQuery}(#1,#2,#3,#4,#5,#6)}
\newcommand{\restrictedRangeQueryOpName}[0]{\mathit{rangeQuery}}

\newcommand{\restrictedRangeQueryOpNew}[8]{\mathit{rangeQuery(#1,#2,#3,#4,#5,#6,#7,#8)}}
\newcommand{\valuesRangeQueryOpNew}[2]{\mathit{valuesRangeQuery(#1,#2)}}
\newcommand{\getCellKOpName}[0]{\mathit{getCell}}
\newcommand{\rangeKOpName}[0]{\mathit{generalRange}}

\mathchardef\mhyphen="2D

\begin{document}

\title
{\large
\textbf{A Compact Representation of Raster Time Series}
\thanks{Funded in part by European Union's Horizon 2020 research and innovation programme under the Marie Sk\l odowska-Curie grant agreement 690941, by Millennium Institute for Foundational Research on Data (IMFD), by Fondecyt-Conicyt grant number 1170497, and by CYTED 519RT0579. Gilberto Guti\'errez was supported by the research group ALgoritmos y BAses de Datos (ALBA) 160119 GI/EF and by the research project 171319 4/R, both funded by Universidad del B\'io-B\'io (Chile).}}

\author{%
Nataly Cruces$^{\ast}$, Diego Seco$^{\ast\dag}$, and Gilberto Guti\'errez$^{\ddag}$\\[0.5em]
{\small\begin{minipage}{\linewidth}
\begin{tabular}{ccc}
$^{\ast}$Universidad de Concepci\'on & $^{\dag}$IMFD - Millennium Institute & $^{\ddag}$Universidad del B\'io-B\'io \\
Concepci\'on, Chile &for Foundational Research& Chill\'an, Chile\\
 \scriptsize{\url{{natalycruces,dseco}@udec.cl}}& on Data, Chile& \scriptsize{\url{ggutierr@ubiobio.cl}}
\end{tabular}
\end{minipage}}
}

\maketitle
\thispagestyle{empty}

\begin{abstract}
\end{abstract}
The raster model is widely used in Geographic Information Systems to represent data that vary continuously in space, such as temperatures, precipitations, elevation, among other spatial attributes. In applications like weather forecast systems, not just a single raster, but a sequence of rasters covering the same region at different timestamps, known as a raster time series, needs to be stored and queried. Compact data structures have proven successful to provide space-efficient representations of rasters with query capabilities. Hence, a naive approach to save space is to use such a representation for each raster in a time series. However, in this paper we show that it is possible to take advantage of the temporal locality that exists in a raster time series to reduce the space necessary to store it while keeping competitive query times for several types of queries.   

\Section{Introduction}

Geographic Information Systems (GIS)~\cite{GIS} facilitate the management of spatial data in digital format, which allow representing features of a certain surface at a time instant, such as rainfall, temperature, population density, among others. Vector and raster are the two main models used in GIS, each of them tailored for different applications. The latter consists in dividing a surface in a regular set of cells, forming a two-dimensional matrix with numeric values that represent some feature of such surface at a given timestamp (e.g. temperature in Snowbird at 2019-03-26 9 am).

The raster model allows representing large areas of land, thus generating huge spatial data volumes. Two factors affect the space needed to store a raster. One is the spatial resolution, the smaller the cell size of a raster, the higher the precision of the model, but also the larger the space consumption. In several domains, not just a single raster, but a sequence of rasters covering the same region at different timestamps has to be stored and queried. This sequence is known in the literature as a raster time series and it is used, for example, in weather forecast systems and in data mining on \emph{Satellite Image Time Series}~\cite{Petitjean2010}. The temporal resolution, defined as the distance in time between two consecutive rasters in the series also impacts the space consumption. For example, going up from a precision of days to a precision of hours requires 24 times more space.

Data compression has been used to reduce the storage and transmission time of raster data~\cite{Kou:1995:DIC:527032,Wallace:1991:JSP:103085.103089}. Even the widely used format to represent rasters, GeoTIFF, supports compression techniques based on Lempel-Ziv-Welch~\cite{Salomon:2006:DCC:1196474}. Recently, compact data structures~\cite{Navarrobook} that not only reduce the space, but also provide index capabilities on the rasters, have been proposed~\cite{BernardoRoca13,LadraPS17,PintoSG17}. Although they do not reduce the space as much as compression techniques, they support several interesting operations without decompression, such as the retrieval of a specific zone or filtering the cells in a zone which values are restricted to a range. These operations are useful in many domains, for example, to zoom-in to a specific area or to detect zones with high flood risk.

A naive approach to store and query raster time series is to use one of the aforementioned compact data structures for each raster in the series. However, in~\cite{PintoSG17} we hypothesized that more compact solutions were possible because of the temporal locality existing in raster time series. Temporal locality produces that the same cell, which represents a spatial region, has similar values in consecutive rasters. For example, the temperature in Snowbird at 14 pm and at 15 pm is often similar. This is even more evident if we increase the temporal resolution, for example, to blocks of 15 minutes. In this paper, we propose a compact data structure that validates the hypothesis in~\cite{PintoSG17}. Our solution is based on the 3D2D mapping described in~\cite{PintoSG17} and a $k^3$-tree~\cite{BernardoPhD}, which is a 3D version of the $k^2$-tree~\cite{Brisaboa14}.

\Section{Background and Related Work}

We refer to~\cite{Navarrobook} for an overview of compact data structures and to the SDSL library~\cite{gbmp2014sea} for practical implementations of many of them. Our proposal is based on a 3D version of the $k^2$-tree~\cite{Brisaboa14}, called the $k^3$-tree~\cite{BernardoPhD}. This data structure has been used in several domains, being the most related to our work the representation of time-evolving region data, a.k.a. moving regions. The {$k^2$-tree} is a compact representation of a quadtree~\cite{samet2006foundations} that stores a sparse binary matrix in compact space by subdividing it into $k^2$ submatrices, which are processed in Morton order storing a 1-bit for non-empty submatrices and a 0-bit, otherwise. This procedure is recursive, but just on non-empty submatrices and up to a minimum submatrix size. This partitioning strategy is generalized by the $k^3$-tree, for 3D, and by the $k^n$-tree, in general. It is important to notice that each 1-bit in the compact representation produces a new (implicit) node with $k^n$-bits, hence degrading compactness with the dimensionality. In a nutshell, both the $k^2$-tree and the $k^3$-tree use few space when the ones in the binary grid are clustered, but non-clustered ones have a more negative impact in the $k^3$-tree than in the $k^2$-tree. This is one of the reasons why there are more examples of success for the $k^2$-tree than for its 3D version.

Our other main building block is one of the compact data structures for rasters presented in~\cite{PintoSG17}, specifically the 3D2D-mapping. This mapping is based on space-filling curves, which are mathematical functions providing a mapping from a multidimensional space to one dimension~\cite{FillingCurve}. The 3D2D-mapping uses the Morton or Z-order~\cite{Morton} curve, which preserves spatial locality and it is suitable for efficient computations based on bit interleaving. Next, we briefly describe this data structure.

Let $<x,y,z>$ be a 3D tuple, with $(x,y)$ a cell in the raster and $z$ its value. These tuples are mapped into a 2D binary grid, in which one axis represents Morton order and the other represents values. Hence, a tuple $<x,y,z>$ induces a 1-bit at row $Z(x,y)$ and column $z$, where $Z(x,y)$ represents the Morton code of cell $(x,y)$. Then, this binary grid is stored using a $k^2$-tree, which exploits the clustering of the 1-bits that is produced by the locality of the Morton order. Three types of queries, $\getCellOpName$, $\rangeQueryOpName$ and $\restrictedRangeQueryOpName$, are supported by this data structure by reduction to primitives of the $k^2$-tree, mainly range queries. However, some of them also require a decomposition of a query into maximal quadboxes~\cite{Proietti1999,Tsai:2004}.

Besides the 3D2D-mapping there are some other compact representations of rasters. For example, in the same paper~\cite{PintoSG17}, another structure based on space-filling curves and succinct trees is proposed, which usually achieves slightly better performance than the 3D2D-mapping, however, it is generalization to raster time series is not easy. Previously, in~\cite{BernardoRoca13} several variants of the $k^2$-tree~\cite{Brisaboa14} were also proposed and proved successful to outperform classical approaches such as GeoTIFF. A more recent work~\cite{LadraPS17} proposed the $k^2$-raster, which is based on an augmented $k^2$-tree that scales better with the number of different values in the raster. Earlier this year~\cite{Cerdeira-PenaBF18}, a generalization of such structure to temporal rasters was also proposed.

\Section{Our solution}

In this section we describe our compact representation of raster time series with support for the following operations:
\begin{itemize}
    \item $\getCellOpNew{x}{y}{t}$ retrieves the value of the cell $(x,y)$ at $t$-th time instant. E.g. temperature in Snowbird\footnote{In our examples, we use a toponym for easy readability, but queries actually receive geographic coordinates as parameters, either to define a point or a window.} at 2019-03-26 9 am.
    \item $\rangeQueryOpNew{x_1}{y_1}{x_2}{y_2}{t_1}{t_2}$ retrieves all the values (a 3D submatrix) registered in a rectangular zone defined by its corners $(x_1,y_1)$ and $(x_2,y_2)$ during a time interval $[t_1,t_2]$. E.g. temperatures in Utah from march to may.
    \item $\restrictedRangeQueryOpNew{x_1}{y_1}{x_2}{y_2}{t_1}{t_2}{\mbox{\it rMin}}{\mbox{\it rMax}}$ retrieves all the cells contained in a rectangular zone defined by $(x_1,y_1)$ and $(x_2,y_2)$ during a time interval $[t_1,t_2]$ which values are in the range $[\mbox{\it rMin},\mbox{\it rMax}]$. E.g. zones (cells) in Utah from march to may with \emph{moderate rain}, i.e. greater than 0.5 mm per hour, but less than 4.0 mm per hour.
    \item $\valuesRangeQueryOpNew{\mbox{\it rMin}}{\mbox{\it rMax}}$ retrieves all the cells which values are in the range $[\mbox{\it rMin},\mbox{\it rMax}]$. Although this is a particular case of the previous type, we study it independently given its application in several domains, for example, to detect zones with high flood risk. 
\end{itemize}

\begin{figure}[h]
\includegraphics[width=\linewidth]{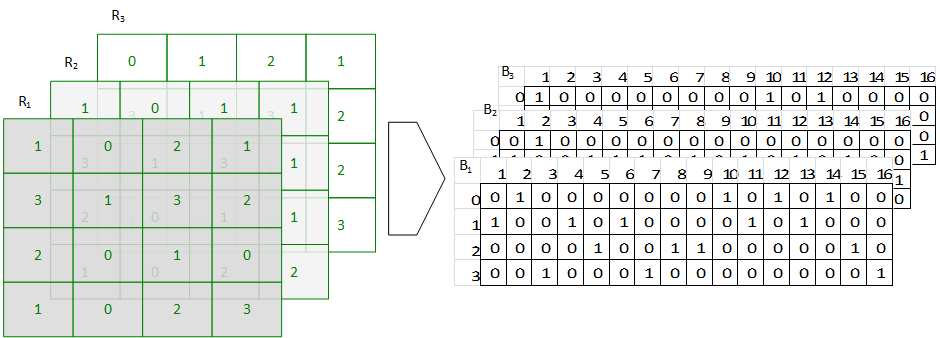}
\caption{Mappings to construct the data structure}\label{fig:example}
\end{figure}

As we mentioned above, our solution is based on the 3D2D-mapping proposed in~\cite{PintoSG17}. Let $TS=[R_1,R_2,\ldots,R_n]$ be a raster time series composed of $n$ rasters. We first read each raster $R_i$ in Morton order, which produces a sequence $T_i$. Then, we transform each $T_i$ to a binary grid $B_i$ that contains a 1-bit at cell $B_i[x][y]$ if $T_i[x]=y$ and a 0-bit, otherwise. Finally, we concatenate all these binary grids into a 3D cube $\{B_1,B_2,\ldots,B_n\}$ and store it using a $k^3$-tree. Figure~\ref{fig:example} shows an example of the mappings used by our method.

In the example, let us focus on the first raster $R_1$, which is transformed into the binary grid $B_1$. The first four values of $R_1$ in Morton order are $1$, $0$, $3$, and $1$, which induce a 1-bit at row $1$ of the first column, at row $0$ of the second column, at row $3$ of the third column, and at row $1$ of the fourth column, respectively. All the other bits in these four columns are 0-bits. Similarly, the first value in $R_2$ is also a 1, which induces a 1-bit at $B_2[1][1]$.

There are two important characteristics in the domain for the $k^3$-tree to perform well, which are the spatial and the temporal locality. Spatial locality is due to Tobler's first law of geography~\cite{Tobler} \emph{``everything is related to everything else, but near things are more related than distant things''}, and to the use of space-filling curves in the mapping, which preserve such property. In other words, Tobler's law implies that cells that are close in a raster of the time series should contain similar values. In our transformation to a binary grid, we use Morton order, and thus cells that are close in the original raster are also close in one of the axis of the grid (w.l.o.g. let us say x-axis). For each of such cells, we set a 1-bit in the row corresponding to its value. As their values are similar, these 1-bits will be close in the binary grid. Hence, the 1-bits in each binary grid are clustered. 

Temporal locality is not related to a specific raster, but to consecutive rasters. This property states that if $R_i[x][y]=v$ and $R_{i+1}[x][y]=v'$, then $v$ and $v'$ should be similar. Given our mapping, both $B_i$ and $B_{i+1}$ will have a 1-bit set in column $Z(x,y)$, and row $v$ and $v'$, respectively. Hence, the 1-bits are not only clustered in each binary grid, but also through consecutive binary grids. The temporal resolution of the raster time series has a great impact on this property, in most domains. This can be easily explained with the following example. Consider two temporal resolutions, one of 1-hour blocks and another of 6-hours blocks, and the domain of temperatures. Obviously, the temperature in Snowbird at 9 am is similar to the temperature at 10 am, but not necessarily similar to the temperature at 3 pm. In the next section, we illustrate this influence with some experiments.

To complete the description of our proposal, we explain the query algorithms to solve the four types of queries explained above. The $k^3$-tree supports operations $\getCellKOpName$, which checks the value of a cell, and $\rangeKOpName$, which performs a 3-dimensional range query, as explained in~\cite[Section 6.2.1]{BernardoPhD}. Then, $\getCellOpName$ can be easily implemented using $\getCellKOpName$, and all the other queries using $\rangeKOpName$. An important consideration is about the spatial dimension of the queries which, due to the use of Morton order, can not be directly transformed to a contiguous range in one dimension of the $k^3$-tree. However, a maximal quadbox decomposition~\cite{Proietti1999,Tsai:2004} can be used to convert a general spatial query, to a set of maximal quadboxes, and each of them becomes a contiguous range in the $k^3$-tree, hence allowing the use of $\rangeKOpName$. This is the same solution used in~\cite{PintoSG17}.

\Section{Experiments}\label{sec:exp}

In this section, we present our preliminary experiments that show the viability of the proposal. We compare the solution described above, with a baseline based on the use of the 3D2D-mapping for each raster of the time series using a $k^2$-tree. This baseline is a representative of the compact data structures existing for rasters~\cite{BernardoRoca13,LadraPS17,PintoSG17}, which exploit the spatial locality but not the temporal one. All the experiments presented here were performed in an Intel Core i7-3820@3.60GHz, 32GB RAM, running Ubuntu server (kernel 3.13.0-35). We compiled with gnu/g++ version 4.6.3 using -O3 directive as both the baseline and the proposed data structure were implemented in C++.

Regarding the datasets, we created several real datasets of temperatures from the data provided by the National Weather Service~\cite{ncep}. Among other products, this service provides hourly information of temperatures for Guam, Hawaii and Puerto Rico. As these data are just available for 48 hours, we downloaded them for two months from September to October 2017. Also, we cropped some parts of these datasets to experiment with datasets of different sizes. Finally, to evaluate the precision of the temporal resolution, we averaged the daily temperatures of the Guam dataset. Table~\ref{tab:datasets} summarizes the characteristics of all the datasets we collected. Also, to illustrate how data vary with time, we prepared a video for two days of the Guam dataset (\url{http://www.inf.udec.cl/~dseco/guam.mp4}).

\begin{table}[!ht]
    \centering
    \begin{tabular}{lccrc}
    \toprule
    Dataset & Columns & Rows & \#Rasters & Resolution\\
    \midrule
    
         Guam-64 & 64 & 64 & 456 & 3-hours\\
         Hawaii-64 &64 & 64 & 1,368 & Hourly \\
         Puerto-Rico-64 &64 & 64 & 1,368 & Hourly \\
         Guam-128 & 128 & 128 & 456 & 3-hours\\
         Hawaii-128 &128 & 128 & 1,368 & Hourly \\
         Puerto-Rico-128 &128 & 128 & 1,368 & Hourly \\
         Guam-193 & 193 & 193 & 456 & 3-hours\\
         Guam-daily-193 & 193 & 193 & 62 & Daily\\
    \bottomrule
    \end{tabular}
    \caption{Characteristics of the datasets}
    \label{tab:datasets}
\end{table}

We first present a space comparison of our proposal and the baseline, which is summarized in Table~\ref{tab:space}. The first column shows the size of the dataset. We show both total space in MB, including the size of the original dataset, and the bits per cell required by both our proposal and the baseline.

\begin{table}[!h]
    \centering
    \begin{tabular}{lrcccc}
    \toprule
    \multirow{2}[2]{*}{Dataset} & \multicolumn{3}{c}{Total space (MB)} & \multicolumn{2}{c}{Bits per cell}\\
    & plain & baseline & proposal & baseline & proposal \\
    \midrule
        Guam-64 & 5.44 & 0.65 &	0.17 & 2.9 & 0.8 \\
        Hawaii-64 & 16.33 & 2.90	& 1.12 &  4.3 &  1.7 \\
        Puerto-Rico-64 & 16.34 & 4.70 & 2.20  &  7.0 &  3.3 \\
        Guam-128 & 21.50 & 1.33 & 0.65 &  1.5  &  0.7\\
        Hawaii-128 & 64.50 & 7.24 & 4.01 &  2.7 &  1.5 \\
        Puerto-Rico-128 & 64.52 & 8.29 & 4.82  & 3.1 &  1.8 \\
        Guam-193 & 48.77 & 2.46	& 2.09  &  1.6  &  1.4 \\
        Guam-daily-193 & 5.37 & 0.75 & 0.81  &  2.7 &  2.9 \\
    \bottomrule
    \end{tabular}
    \caption{Space evaluation}
    \label{tab:space}
\end{table}

The main conclusion of this evaluation is that our proposal usually requires less than half the space of the baseline, thus validating the hypothesis that it is possible to exploit the temporal locality in raster time series to design more space-efficient data structures for them. A second conclusion is the influence of the temporal resolution in our proposal, which can be observed in the last two rows. The original Guam dataset has a temporal resolution of blocks of 3 hours, unlike the others that provide an hourly resolution. Despite this lower resolution, our proposal still slightly outperforms the baseline in this dataset. However, in the last dataset, we averaged the temperatures of each day, generating a dataset with daily resolution. As it can be seen in the last row, our proposal requires slightly more space than the baseline in this case, which is due to the lower temporal locality.

\begin{figure}[!h]
\centering 
\epsfig{file=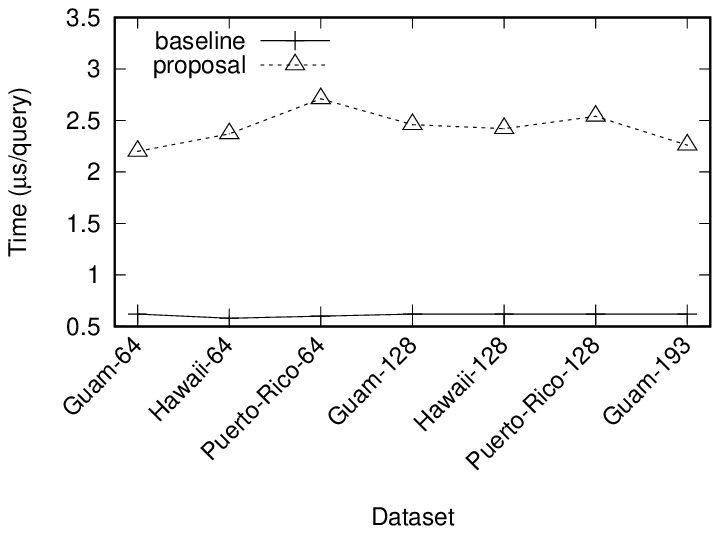} 
\caption{Query time for $\getCellOpName$}\label{fig:access}
\end{figure}

We now present an evaluation of the query time performance for the queries introduced above. First, Figure~\ref{fig:access} shows the results for $\getCellOpName$. We show average time per query for a set of 10,000 random queries.

Results are consistent through the different datasets and show that the proposal is about 4 times slower than the baseline, but still in the order of few microseconds per query. This is not a surprising result as the underlying data structure of the proposal is more complex than that of the baseline, and this type of query just involves one raster in the baseline.

\begin{figure}[!h]
\centering 
\epsfig{file=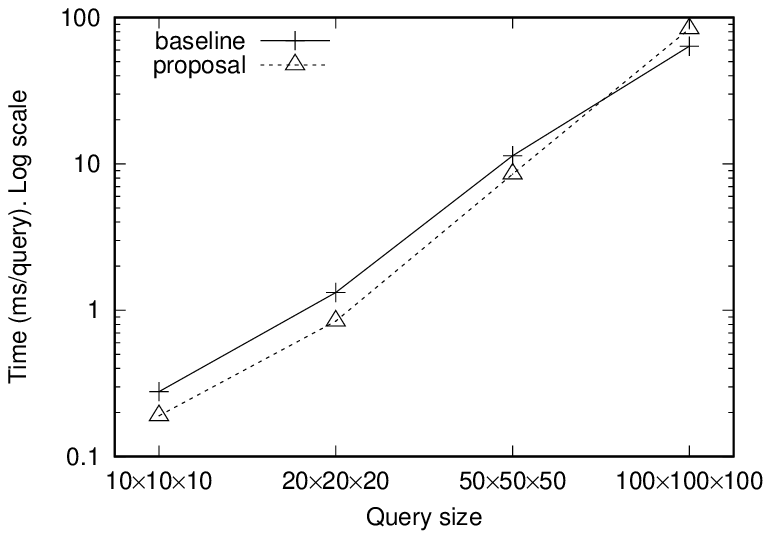} 
\caption{Query time for $\rangeQueryOpName$}\label{fig:window}
\end{figure}

For the other types of queries, we just show the results for one of the datasets, as results are similar in all of them and, in this way, we can focus on the influence of the parameters of such query types. We selected Hawaii-128 as a representative dataset. Recall that this dataset has a spatial extent of $128\times128$ with $1,368$ snapshots and an hourly temporal resolution. Figure~\ref{fig:window} illustrates the results of $\rangeQueryOpName$ for different window sizes. These results show that the proposal is also competitive for this type of queries as the differences between the proposal and the baseline are not significant, being our proposal slightly faster than the baseline for most window sizes (note the logarithmic scale).

\begin{figure}[!h]
\centering 
\epsfig{file=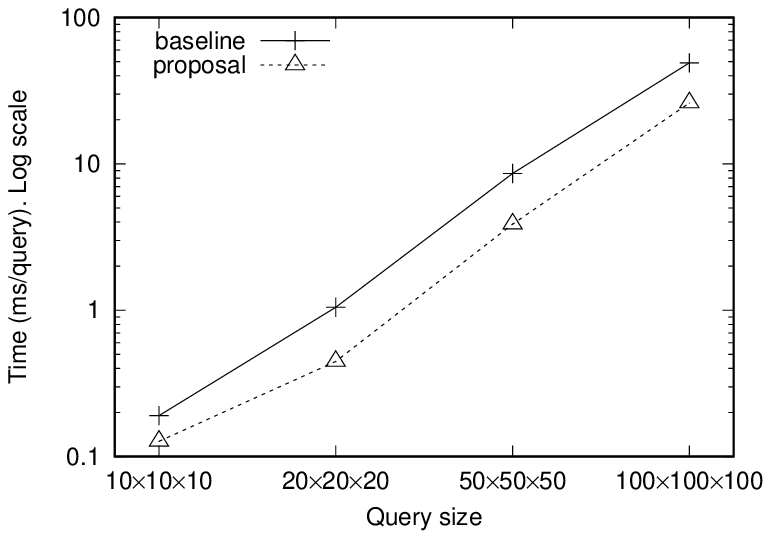} 
\caption{Query time for $\restrictedRangeQueryOpName$, given a fixed range size of $10$}\label{fig:range_r10}
\end{figure}

\begin{figure}[!h]
\centering 
\epsfig{file=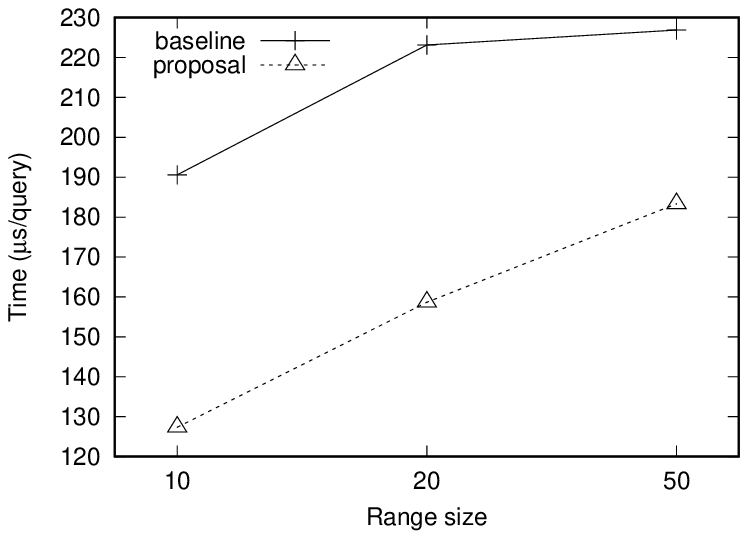} 
\caption{Query time for $\restrictedRangeQueryOpName$, given a fixed window size of $10\times10\times10$}\label{fig:range_w10}
\end{figure}

For the most general type of queries, $\restrictedRangeQueryOpName$, we show the influence of both the query size, given a fixed range size of $10$, and the influence of the range size, given a fixed size of query of $10\times10\times10$. Figures~\ref{fig:range_r10} and~\ref{fig:range_w10} show these results.

Similar to previous experiments, the proposal is slightly faster than the baseline. Another conclusion is that the range size has less influence in the query time than the window size (note the logarithmic scale in Figure~\ref{fig:range_r10}). This is not only because range size represents just one of the dimensions of the $k^3$-tree, but also because it does not require the maximal quadbox decomposition needed for the spatial dimension. 

\Section{Conclusions}

We proposed a new space-efficient data structure that takes advantage of the spatial and temporal locality existing in a raster time series. Our experimental evaluation, performed using a real dataset of temperatures that has both spatial and temporal locality, shows that our structure requires half the space of a baseline that uses a compact representation for each raster in the time series but does not take advantage of the temporal locality in the time series. However, we also showed that our proposal is dependant on the temporal resolution and, in the case of temperatures, it requires similar space than the baseline for daily averaged datasets. Regarding query times, our structure is competitive, and even slightly faster, than the baseline for several types of queries. As future work, a more detailed experimental evaluation is necessary to compare with all the compact data structures for rasters existing in the literature~\cite{BernardoRoca13,LadraPS17,PintoSG17}. Also, our work has to be evaluated against a recent proposal for temporal rasters presented in~\cite{Cerdeira-PenaBF18}. Finally, to deal with the curse of dimensionality of the $k^3$-tree a 3D version of the heavy path tree in~\cite{GagieGLNS15} may improve the space usage.

\Section{References}
\bibliographystyle{IEEEtran}
\bibliography{refs}

\end{document}